\documentclass[aps,prc,twocolumn,superscriptaddress,showpacs]{revtex4}
\usepackage{bm}

\newcommand{\gvct}[1]{\bm{#1}}
\newcommand{\vct}[1]{{\bf #1}}
\newcommand{\ket}[1]{|#1\rangle}
\newcommand{\mtrix}[3]{\langle \,#1\,|\,#2\,|\,#3\,\rangle}
\newcommand{\fslash}[1]{{\ooalign{\hfil/\hfil\crcr$#1$}}}
\newcommand{\kf}{k_{\text{F}}}
\newcommand{\ekf}{E_{\text{F}}}
\newcommand{\kfp}{k_{\text{p}}}
\newcommand{\kfn}{k_{\text{n}}}
\newcommand{\vf}{v_{\text{F}}}
\newcommand{\mitg}{{\mit\Gamma}}
\newcommand{\thp}[1]{\theta^{\text{(p)}}_{\vct{#1}}}
\newcommand{\thn}[1]{\theta^{\text{(n)}}_{\vct{#1}}}
\newcommand{\ep}[1]{E_{\vct{#1}}}
\newcommand{\ekfp}{E_{\text{Fp}}}
\newcommand{\ekfn}{E_{\text{Fn}}}
\newcommand{\ekfi}{E_{\text{Fi}}}

\begin{document}

\title{The Energy-Weighted and Non Energy-Weighted Gamow-Teller
Sum Rules in Relativistic Random  Phase Approximation}

\author{Haruki Kurasawa}
\affiliation{
Department of Physics, Faculty of Science, Chiba University, Chiba
263-8522, Japan
}

\author{Toshio Suzuki}
\affiliation{
Department of Applied Physics, Fukui University, Fukui 910-8507, Japan\\
and\\
RIKEN, 2-1 Hirosawa, Wako-shi, Saitama 351-0198, Japan
}

\begin{abstract}
The non energy-weighted Gamow-Teller(GT) sum rule is satisfied in
relativistic models, when all nuclear density-dependent terms,
including Pauli blocking terms from nucleon-antinucleon excitations,
are taken into account in the RPA correlation function. The no-sea
approximation is equivalent to this approximation for the giant GT
resonance state and satisfies the sum rule, but each of the total
$\beta_-$ and $\beta_+$ strengths is different in the two
approximations. It is also shown that the energy-weighted sum of
the GT strengths for the $\beta_-$ and $\beta_+$ transitions in
RPA is equal to the expectation value of the double commutator
of the nuclear Hamiltonian with the GT operator, when the
expectation value is calculated with the ground state in the mean
field approximation. Since the present RPA neglects renormalization
of the divergence, however, the energy-weighted strengths outside
of the giant GT resonance region become negative. These facts are
shown by calculating in an analytic way the GT strengths of
nuclear matter.
\end{abstract}

%% \keywords{Gamow-Teller states}

\maketitle

\section{Introduction}

Recently, the present authors have investigated Ikeda-Fujii-Fujita(IFF)
sum rule\cite{iff} of Gamow-Teller(GT)
transitions in relativistic models\cite{ksg1,ksg1-2}. It has been
shown that the sum of the GT strengths in the nucleon sector
is quenched by about 6\% in finite nuclei\cite{ksg1-2} and 12\%
in nuclear matter\cite{ksg1}.
This fact is  owing to the small component of nucleon wave functions,
and the quenched amount is taken by 
nucleon-antinucleon states.
Although the nucleon-antinucleon states are far from the giant GT
resonance region, if there is a coupling with particle-hole states,
they may contribute to nuclear excitations
in the giant resonance region as virtual states.
In the previous paper\cite{ksg1,ksg1-2}, the effects of
the coupling were estimated with use of the
random phase approximation(RPA) in relativistic nuclear models.
RPA correlation functions were calculated by neglecting
the nuclear density-independent terms in order to avoid the divergence,
but by keeping
all the density-dependent terms
including the Pauli blocking ones
from nucleon-antinucleon excitations.
This method was frequently used in the study of no charge-exchange
excitations\cite{chin,ks1,ks2}. 
From now on let us call this approximation no free term
approximation(NFA). In this calculation, we found that effects of
the coupling are negligible on the excitation
energy and strength of the GT state in the nucleon sector.

The purpose of the present paper is threefold. First, we will study
whether or not NFA satisfies IFF sum rule which is
the non energy-weighted sum rule for
the difference between the $\beta_-$ and $\beta_+$
transition strengths. When we discuss excitation strengths,
the used approximation should satisfy at least
the model-independent sum rule.
It seems not to be trivial for NFA to satisfy IFF sum rule, since
for the $\beta_- $ transitions, the backward amplitudes
in the particle-hole sector do not contribute to the
strength as in Tamm-Dankoff approximation,
while in the nucleon-antinucleon excitations, they
are comparable to the forward ones in the present model.
We will show, however, 
that the sum rule is satisfied in NFA.
The proof will be done in an analytic way for nuclear matter.

Second, we will discuss the sum rule in the no-sea approximation(NSA),
which is also used frequently in no charge-exchange
excitations\cite{furun,ma}.
It will be shown that NSA also
satisfies IFF sum rule, but that each total strength of
the $\beta_-$ and $\beta_+$ transitions 
is different from those of NFA.
As far as the giant GT resonance state, however, its
excitation energy and strength are predicted in the same way
as in NFA.

Third, we will investigate the energy-weighted sum of the GT strengths
in RPA, which have not been discussed so far, as far as the authors
know. For no charge-exchange excitations, there is a famous theorem
in the non-relativistic framework
that the energy-weighted sum of the excitation strengths in RPA
is equal to the expectation value of the double commutator
of the nuclear Hamiltonian with
the relevant operator\cite{thou}.
Here the expectation value is calculated with
the Hartree-Fock ground state.
It will be shown in relativistic models that the same theorem
holds for charge-exchange excitations with respect to
the sum of the GT strengths for $\beta_-$ and $\beta_+$ transitions.
In both NFA and NSA, however, the energy-weighted strength outside
of the giant GT resonance region is negative.
This is due to the fact that divergent terms are simply
neglected in the two approximations. We definitely need the
renormalization to solve this problem in future work.
    
We note here that
such detailed discussions as mentioned in the above are required
not only from a theoretical point of view, but also from
the recent experiment.  It has been shown experimentally
that the sum rule value is quenched by about 10\% from the
non-relativistic analysis of $(p,n)$ reaction\cite{wakasa}.
So far, all the 10\% quenching was assumed to be 
due to the coupling of the particle-hole states with $\Delta$-hole
states\cite{sakai,bentz}.
If the 6\% quenching is due to the relativistic effects,
it may conclude that only 4\% stems from the contribution of the
$\Delta$ degrees of freedom. Such a weak coupling of
the particle-hole states with the $\Delta$-hole states may
yield the extremely small value of Landau-Migdal parameter
$g'_{\text{N}\Delta}$ which dominates the critical density of
the pion condensation\cite{ksg2} and other spin-dependent
response functions\cite{ichimura}.
Thus we need to discuss GT strengths carefully, before studying
spin-dependent structure of nuclei in detail.

In the next section, we will briefly review NFA and NSA in no
charge-exchange excitations. In section \ref{sec3}, the mean field
correlation function of the GT excitations will be derived.
In section \ref{sec4}, we will discuss the GT sum rules in RPA
within the framework of NFA and NSA. The final section will be
devoted to discussions and conclusions of the present paper.

\section{NFA and NSA} \label{sec2}

The fundamental problem in relativistic nuclear
models is how to renormalize the divergence
due to the anti-nucleon degrees of freedom. Only few attempts
of the renormalization have been reported so far\cite{chin, ks3}.
Nevertheless, it has been shown for the past 30 years that
relativistic models explain phenomenologically very well 
nuclear structure
and reactions without renormalization\cite{sw,ring}.
In those studies, there are some cases where it is necessary
to take into account
a part of the antinucleon degrees of freedom.
In RPA based on the mean field approximation,
the continuity equation of the baryon current is violated,
if the configuration space is limited to the nucleon sector only. 
At present, there are two ways to avoid the violation.
The one(NFA) is to neglect the
density-independent terms in the RPA correlation function
which are divergent, but to keep all
the density-dependent terms including the Pauli blocking terms
from nucleon-antinucleon excitations\cite{chin,ks1,ks2}.
The other is called the no-sea approximation(NSA), where
the Dirac sea is assumed to be empty, and the antinucleon states
are treated as particle states with negative energies
in the configuration space of RPA\cite{furun}. NSA is
described by changing the sign of the imaginary part of the Green
function of the antinucleon. It is shown in this way that the
divergence of the RPA correlation function disappears and that
the continuity equation is not violated\cite{furun}.

In this section, we briefly review the above two approximations in
nuclear matter, since
their structure in charge-exchange excitations is a little different
from that in no charge-exchange ones discussed so far.

The Green function of the Dirac particle in nuclear matter
is written as\cite{chin,sw}, 
\begin{eqnarray}
G(p)=\left(1 -\theta_{\vct{p}}\right)G_{\text{p}}(p)
 + \theta_{\vct{p}}\,G_{\text{h}}(p)
 + G_{\bar{\text{N}}}(p),\label{green}
\end{eqnarray}
where the particle, hole and antinucleon parts are given,
respectively, as
\begin{eqnarray}
G_{\text{p}}(p)
&=&\frac{{\mit\Lambda}_+(\vct{p})}{p_0-E_{\vct{p}}+i\varepsilon}\,,\\
G_{\text{h}}(p)
&=&\frac{{\mit\Lambda}_+(\vct{p})}{p_0-E_{\vct{p}}-i\varepsilon}\,,\\
G_{\bar{\text{N}}}(p)
&=&
-\,\frac{{\mit\Lambda}_-(\vct{p})}{p_0+E_{\vct{p}}-i\varepsilon}.
\label{nbar}
\end{eqnarray}
The projection operators in the above equation are defined as usual,
\begin{eqnarray*}
{\mit\Lambda}_+(\vct{p})
&=&\sum_\alpha u_\alpha(\vct{p}) \overline{u}_\alpha(\vct{p})
=\frac{\fslash{p}+M^\ast}{2E_{\vct{p}}}\\
{\mit\Lambda}_-(\vct{p})
&=&-\sum_\alpha v_\alpha(-\vct{p}) \overline{v}_\alpha(-\vct{p})
=\frac{\fslash{\tilde{p}}+M^\ast}{2E_{\vct{p}}}      
\end{eqnarray*}
with
\[
p^\mu=(E_{\vct{p}},\vct{p})\,,\quad
\tilde{p}^\mu=(-\,E_{\vct{p}},\vct{p})\,,\quad
E_{\vct{p}}=\sqrt{M^{\ast 2}+\vct{p}^2}.
\] 
The step function is expressed by using the abbreviation:
$\theta_{\vct{p}}=\theta(\kf-|\vct{p}|)$, $\kf$ being the Fermi
momentum. The Green function Eq.(\ref{green}) can be also
expressed in terms of the density-dependent and -independent
parts as, 
\begin{eqnarray}
G(p)&=&G_{\text{D}}(p)+G_{\text{F}}(p) \label{green2}\,,\\
G_{\text{D}}(p)
&=&\theta_{\vct{p}}
\left( G_{\text{h}}(p)- G_{\text{p}}(p)\right)\,,\\
G_{\text{F}}(p)
&=&G_{\text{p}}(p)+G_{\bar{\text{N}}}(p).
\end{eqnarray}

The mean field correlation function is described as
\begin{eqnarray}
{\mit\Pi}&=&-\,\frac{1}{2\pi i}
\int\!d^4p\,\text{Tr}
\left( \gamma_a G(p+q)\gamma_b G(p) \right), \label{hartree}
\end{eqnarray}
where $\gamma_a$ and $\gamma_b$ stand for the $4\times 4$ matrix
of the external field. 
If Eq.(\ref{green2}) is inserted into Eq.(\ref{hartree}),
the mean field correlation
function is divergent, because of the density-independent term which
contains $G_{\text{F}}G_{\text{F}}$. In NFA, it is simply neglected
without renormalization, to have
\begin{widetext}
\begin{eqnarray}
{\mit\Pi}_{\text{D}}&=&-\,\frac{1}{2\pi i}
\int\!d^4p\,\text{Tr}\,\gamma_a
\left(
 G_{\text{D}}(p+q)\gamma_b G_{\text{D}}(p)
 +G_{\text{F}}(p+q)\gamma_b G_{\text{D}}(p)
 +G_{\text{D}}(p+q)\gamma_b G_{\text{F}}(p)
 \right).
\label{nfahartree}
\end{eqnarray}
Keeping the terms which remain after integration over $p_0$ in
the complex plane,
the above equation is separated into two parts,
\begin{eqnarray}
{\mit\Pi}_{\text{D}}&=&
{\mit\Pi}_{\text{ph}}+{\mit\Pi}_{\text{Pauli}}.
\end{eqnarray}
The first term represents the particle-hole correlation function,
while the second term the nucleon-antinucleon one, 
\begin{eqnarray}
{\mit\Pi}_{\text{ph}}&=&
 -\,\frac{1}{2\pi i}
 \int\!d^4p\,\text{Tr}\,\gamma_a
 \left(
  (1-\theta_{\vct{p}+\vct{q}})\,\theta_{\vct{p}}
  \,G_{\text{p}}(p+q)\gamma_b G_{\text{h}}(p)% \Bigr. \nonumber \\
%& &
% \phantom{
% -\,\frac{1}{2\pi i}
% \ \ \!d^4p\,\text{Tr}\,\gamma_a 
% }
 +%\,\Bigl.
 (1-\theta_{\vct{p}})\,\theta_{\vct{p}+\vct{q}}
  \,G_{\text{h}}(p+q)\gamma_b G_{\text{p}}(p) \right), \label{p-h} \\
 {\mit\Pi}_{\text{Pauli}}&=&
 \,\frac{1}{2\pi i}
 \int\!d^4p\,\text{Tr}\,\gamma_a
 \left(
  \theta_{\vct{p}+\vct{q}}
  \,G_{\text{p}}(p+q)\gamma_b G_{\bar{\text{N}}}(p)
  +
  \theta_{\vct{p}}
  \,G_{\bar{\text{N}}}(p+q)\gamma_b G_{\text{p}}(p)
 \right). \label{Pauli}
\end{eqnarray}
Thus, ${\mit\Pi}_{\text{Pauli}}$ is composed of the Pauli blocking
terms. In NFA, however, these terms are necessary for keeping
 the continuity equation of
the baryon current, and for satisfying the GT
sum rule, as shown later. 

In NSA, Green function of the antinucleon in Eq.(\ref{nbar}) is
artificially modified by changing a sign of the imaginary part,
\begin{eqnarray}
G_{\text{no}}(p)
=-\,\frac{{\mit\Lambda}_-(\vct{p})}{p_0+E_{\vct{p}}+i\varepsilon}.
\end{eqnarray}
Then density-independent terms 
have no contribution to the integration over
$p_0$ in Eq.(\ref{hartree}), and we can avoid the divergence problem
without violating the continuity equation\cite{furun}.
The terms which should be kept in NSA are written as
\begin{eqnarray}
{\mit\Pi}&=&
 {\mit\Pi}_{\text{ph}}+{\mit\Pi}_{\text{no-sea}},
\end{eqnarray}
where the particle-hole correlation function is the same as in NFA,
but the antinucleon-dependent part is given by
\begin{eqnarray}
 {\mit\Pi}_{\text{no-sea}}&=&
 -\,\frac{1}{2\pi i}
 \int\!d^4p\,\text{Tr}\,\gamma_a
 \left(
  \theta_{\vct{p}+\vct{q}}
  \,G_{\text{h}}(p+q)\gamma_b G_{\text{no}}(p)
  +
  \theta_{\vct{p}}
  \,G_{\text{no}}(p+q)\gamma_b G_{\text{h}}(p)
 \right). \label{no-sea}
\end{eqnarray}

After integration over $p_0$,
the mean field correlation functions in NFA and NSA are
finally written, respectively, as
\begin{eqnarray}
{\mit\Pi}_{\text{ph}}
&=&
\int\!d^3p\,\text{Tr}
\left(
\gamma_a {\mit\Lambda}_+(\vct{p}+\vct{q})
 \gamma_b {\mit\Lambda}_+(\vct{p})
\right)
\left(
\frac{\left(1-\theta_{\vct{p}+\vct{q}}\right)\theta_{\vct{p}}}
{E_{\vct{p}+\vct{q}}-E_{\vct{p}}-q_0-i\varepsilon}
-
\frac{\left(1-\theta_{\vct{p}}\right)\theta_{\vct{p}+\vct{q}}}
{E_{\vct{p}+\vct{q}}-E_{\vct{p}}-q_0+i\varepsilon}
\right), \label{p-h1}\\
\noalign{\vskip8pt}
{\mit\Pi}_{\text{Pauli}}&=&
\int\!d^3p\,\text{Tr}
\left(
 \theta_{\vct{p}+\vct{q}}
 \,
 \frac{\gamma_a{\mit\Lambda}_+(\vct{p}+\vct{q})
 \gamma_b{\mit\Lambda}_-(\vct{p})}
 {E_{\vct{p}+\vct{q}}+E_{\vct{p}}-q_0-i\varepsilon}
 +
 \theta_{\vct{p}}
 \,
 \frac{\gamma_a{\mit\Lambda}_-(\vct{p}+\vct{q})
 \gamma_b{\mit\Lambda}_+(\vct{p})}
 {E_{\vct{p}+\vct{q}}+E_{\vct{p}}+q_0-i\varepsilon}
\right),\label{nfa} \\ 
\noalign{\vskip8pt}
{\mit\Pi}_{\text{no-sea}}
&=&
\int\!d^3p\,\text{Tr}
\left(
 \theta_{\vct{p}+\vct{q}}
 \,
 \frac{\gamma_a{\mit\Lambda}_+(\vct{p}+\vct{q})
 \gamma_b{\mit\Lambda}_-(\vct{p})}
 {E_{\vct{p}+\vct{q}}+E_{\vct{p}}-q_0+i\varepsilon}
 +
 \theta_{\vct{p}}
 \,
 \frac{\gamma_a{\mit\Lambda}_-(\vct{p}+\vct{q})
 \gamma_b{\mit\Lambda}_+(\vct{p})}
 {E_{\vct{p}+\vct{q}}+E_{\vct{p}}+q_0+i\varepsilon}
\right). \label{nsa}
\end{eqnarray} 
\end{widetext}
Thus, the real parts are the same in NFA and NSA, while the imaginary
parts of ${\mit\Pi}_{\text{Pauli}}$ and ${\mit\Pi}_{\text{no-sea}}$
have an opposite sign to each other. This fact implies that both
approximations yield the same excitation energies for the discrete
states in RPA,
but there is a possibility that their response functions and 
sum values of the excitation strengths are different
from each other.
An example will be shown later in the case of the GT strengths. 

We note that the relationship between the Landau-Migdal parameters
and the correlation functions is the same in NFA and NSA, since it
depends on the only real part of the correlation functions,
as shown in ref.\cite{ks2,ks4}. All physical quantities expressed
in terms of the Landau-Migdal parameters, therefore, must be the
same in the two approximations.
The reason why NSA as well as NFA
does not violate the continuity equation is
because it is also independent of the imaginary part.
This fact is verified by showing explicitly
\begin{eqnarray}
{\mit\Pi}_{\text{ph}}+ {\mit\Pi}_{\text{Pauli}}
={\mit\Pi}_{\text{ph}}+ {\mit\Pi}_{\text{no-sea}}
=0
\end{eqnarray}
for the operator $\gamma_b =\,\,\fslash{q}$.
Each of them for $\gamma_b =\,\,\fslash{q}$ is written,
respectively, as\cite{ks2}
\begin{eqnarray*}
{\mit\Pi}_{\text{ph}}
 &=&
 \int\!d^3p\,
 \frac{\theta_{\vct{p}}}
 {4E_{\vct{p}}E_{\vct{p}+\vct{q}}}
 \,\text{Tr}
 \left(
 \gamma_a \gamma_0
 \left(
  \fslash{\tilde{p}}' \fslash{p} - \fslash{p} \fslash{\tilde{p}}'
      \right)
 \right), \\ %% \label{current_ph}\\       
{\mit\Pi}_{\text{Pauli}}
 &=&{\mit\Pi}_{\text{no-sea}}\nonumber \\
 &=&\int\!\!d^3p\,
 \frac{\theta_{\vct{p}}}{4E_{\vct{p}}E_{\vct{p}+\vct{q}}}
 \text{Tr}
 \left(
 \gamma_a \gamma_0
 \left( \fslash{p}\fslash{p}' - \fslash{p}'\fslash{p} \right)
 \right), %%\label{current_nnbar}
\end{eqnarray*}
using the notations:
\[
p'^{\mu}=(\,E_{\vct{p}+\vct{q}}\,,\,\vct{p}+\vct{q}\,)\,,\quad
\tilde{p}'^\mu =(\,-E_{\vct{p}+\vct{q}}\,,\,\vct{p}+\vct{q}\,).
\]

\section{The Mean Field Correlation Functions for GT Excitations}

\label{sec3}

When we discuss
GT excitations in $N\ne Z$ nuclei,
we need the following replacement in Eqs.(\ref{p-h1}), (\ref{nfa})
and (\ref{nsa}),
\begin{eqnarray*}
\gamma_a&\rightarrow& \gamma_a=\gamma_5\gamma_y\,\tau_+\,,\quad 
\gamma_b\rightarrow \gamma_b=\gamma_5\gamma_y\,\tau_-\,,\\
\noalign{\vskip4pt} 
 \theta_{\vct{p}}&\rightarrow&
 \frac{1-\tau_z}{2}\,\theta_{\vct{p}}^{\text{(p)}}
+ \frac{1+\tau_z}{2}\,\theta_{\vct{p}}^{\text{(n)}},
\end{eqnarray*}
where we have defined the isospin operators for the convenience as 
\begin{eqnarray*}
\tau_\pm =\left(\tau_x \pm i\tau_y\right)/\sqrt{2}\, \ \ ,\qquad
\tau_0=\tau_z,
\end{eqnarray*}
and the step function with respect to the proton($k_p$)
and neutron Fermi momentum($k_n$) :
\[
\theta_{\vct{p}}^{(i)}=\theta(k_i -|\vct{p}|)\,,\quad i=p,\, n\,.
\]
Then, for the $\beta_-$ excitation at $\vct{q}=0$ in $N>Z$ nuclei,
Eq.(\ref{p-h1}) becomes to be 
\begin{eqnarray}
{\mit\Pi}_{\text{ph}}
 =-\,4\int\!\!d^3p\,\frac{M^{\ast2}+p_y^2}{\ep{p}^2}
\frac{\thn{p}-\thp{p}}{q_0+i\varepsilon},\label{p-h2}
\end{eqnarray}
and
Eqs.(\ref{nfa}) and (\ref{nsa}) are described, respectively, as
\begin{eqnarray}
 {\mit\Pi}_{\text{Pauli}}&=&
4\int\!d^3p\,\frac{\vct{p}^2-p_y^2}{E_{\vct{p}}^2} \nonumber \\
& &
\times\left(
\frac{\theta_{\vct{p}}^{\text{(p)}}}
{q_0-2E_{\vct{p}}+i\varepsilon}
-\frac{\theta_{\vct{p}}^{\text{(n)}}}
{q_0+2E_{\vct{p}}-i\varepsilon}
\right), \ \ \ \label{pi_pauli2} \\
\noalign{\vskip4pt}
{\mit\Pi}_{\text{no-sea}}
&=&
-\,4\int\!d^3p\,\frac{\vct{p}^2-p_y^2}{E_{\vct{p}}^2}
\nonumber \\
& &
\times \left(
\frac{\theta_{\vct{p}}^{\text{(n)}}}
{q_0+2E_{\vct{p}}+i\varepsilon}
-\frac{\theta_{\vct{p}}^{\text{(p)}}}
{q_0-2E_{\vct{p}}-i\varepsilon} 
\right). \label{pi_nosea2}
\end{eqnarray}
From the above equations, we see, at this stage,
first that the present model has only the forward amplitudes
for particle-hole pairs.
Second, in NFA, the forward amplitudes of the
nucleon-antinucleon pairs have an opposite sign to the one of the
particle-hole pairs. This means that the GT strength from the Pauli
blocking terms is negative. Third, in NSA, the excitation energy of
the antinucleon-hole pairs is negative, as mentioned before.
Finally we can see that the
density dependence of NSA, expressed by the step functions, is
different from the one of NFA. As a result, their GT strengths are
different from each other,
although the difference between the strengths of
$\beta_-$ and $\beta_+$ transitions is the same in the two
approximations.

The correlation functions for the $\beta_+$ transitions are
obtained by changing the sign of $q_0$ in Eqs.(\ref{p-h2}) to
(\ref{pi_nosea2}).
In this case, there are no forward amplitudes for particle-hole
pairs, but backward ones. 

By performing  the integration, Eq.(\ref{p-h2}) is written as
\begin{eqnarray*}
{\mit\Pi}_{\text{ph}}
=-\,\frac{\alpha_{\text{ph}}}{q_0+i\varepsilon}\,,\quad
\alpha_{\text{ph}}
=\frac{16\pi}{3}\left( Q(k_n)-Q(k_p) \right),% \label{p-h3}
\end{eqnarray*}
where $Q(k_i)$ is defined by
\begin{eqnarray}
Q(k_i)&=&\frac{3}{4\pi}\int_0^{k_i}\!d^3p\,
\frac{M^{\ast2}+p_y^2}{\ep{p}^2} \nonumber \\
&=&\frac{k_i^3}{3}+2k_i M^{\ast2}
-2M^{\ast3}\tan^{-1}\frac{k_i}{M^\ast}. \label{qkf}
\end{eqnarray}
Eqs.(\ref{pi_pauli2}) and (\ref{pi_nosea2})
are expressed by separating into the real and 
imaginary parts:
\begin{eqnarray*}
{\mit\Pi}_{\text{Pauli}}
&=&{\mit\Pi}_{\text{Pauli}}^{\text{(R)}}
+i{\mit\Pi}_{\text{Pauli}}^{\text{(I)}} \\
\noalign{\vskip4pt} 
{\mit\Pi}_{\text{no-sea}}
&=&{\mit\Pi}_{\text{no-sea}}^{\text{(R)}}
+i{\mit\Pi}_{\text{no-sea}}^{\text{(I)}},
%\label{nbar2}
\end{eqnarray*}
which are related with each other as
\begin{eqnarray*}
{\mit\Pi}_{\text{Pauli}}^{\text{(R)}}
 ={\mit\Pi}_{\text{no-sea}}^{\text{(R)}}\,,\quad
{\mit\Pi}_{\text{Pauli}}^{\text{(I)}}
=-{\mit\Pi}_{\text{no-sea}}^{\text{(I)}}.
\end{eqnarray*}
The explicit forms of ${\mit\Pi}_{\text{Pauli}}^{\text{(R)}}$ and
${\mit\Pi}_{\text{Pauli}}^{\text{(I)}}$ are described as
\begin{eqnarray*}
{\mit\Pi}_{\text{Pauli}}^{\text{(R)}}
&=&-\,\frac{16\pi}{3}
\left(
  P_{\overline{\text{N}}}(\kfp,-\,q_0)
+ P_{\overline{\text{N}}}(\kfn,q_0)
\right)\,,\\
{\mit\Pi}_{\text{Pauli}}^{\text{(I)}}
&=& I_{\text{Pauli}}(\kfp,q_0)
  + I_{\text{Pauli}}(\kfn,-\,q_0),%\label{r-i}
\end{eqnarray*}
where we have defined the two functions:
\begin{eqnarray*}
P_{\overline{\text{N}}}(\kf,q_0)&=&
\frac{3}{4\pi}\int_0^{\kf}\!\frac{d^3p}{\ep{p}^2}
\frac{\vct{p}^2-p_y^2}{2\ep{p}+q_0}\,,\\
I_{\text{Pauli}}(\kf,q_0)&=&
-\,4\pi\int_0^{\kf}\!\!d^3p\,\frac{\vct{p}^2-p_y^2}{\ep{p}^2}
\,\delta(q_0-2\ep{p}).
\end{eqnarray*}
The second function is simply expressed as
\begin{eqnarray}
& & I_{\text{Pauli}}(\kf,q_0) \nonumber \\
&=&
\left\{
\begin{array}{ll}
\displaystyle{-\,
\frac{4\pi^2}{3}\frac{\left(q_0^2-4M^{\ast2}\right)^{3/2}}{q_0}
}\,, & 2M^\ast<q_0<2\ekf,\ \ \  \label{Ipauli}\\
\noalign{\vskip4pt}
0\,, & \text{otherwise},
\end{array}
\right.
\end{eqnarray}
with $\ekf = \sqrt{M^{\ast2}+\kf^2}$, while the first one is
written as
\begin{eqnarray*}
P_{\overline{\text{N}}}(\kf,q_0)
 &=& \frac{\kf\ekf}{2}-\frac{q_0\kf}{2}
-\frac{6M^{\ast2}-q_0^2}{4}\log\frac{\kf+\ekf}{M^\ast}
\nonumber \\
& &
 +\,\frac{2M^{\ast3}}{q_0}\tan^{-1}\frac{\kf}{M^\ast}
+\frac{4M^{\ast2}-q_0^2}{4q_0}I_{\text{B}},
\end{eqnarray*}
where $I_{\text{B}}$ is given by
\begin{eqnarray*} 
I_{\text{B}}&=&
 \int_{\ekf-\kf}^{M^\ast}\!dt\,
\frac{q_0^2-4M^{\ast2}}{t^2+q_0t+M^{\ast2}} \\
\noalign{\vskip4pt} 
&=&\left\{
\begin{array}{ll}
-\,4M^{\ast}\sqrt{1-x^2}\tan^{-1}y\,, & x^2<1\\
\noalign{\vskip4pt}
\displaystyle{
2M^{\ast}\sqrt{x^2-1}
\log\left| \frac{1+y}{1-y} \right|
}\,, & x^2>1 
\end{array}
\right.
\end{eqnarray*}
with
\[
 y=\frac{\ekf-M^\ast}{\kf}\frac{\sqrt{|1-x^2|}}{1+x}\,,\qquad
 x=\frac{q_0}{2M^\ast}.
\]
For $q_0\ll 2M^\ast$, we have
\begin{eqnarray*}
P_{\overline{\text{N}}}(\kf,q_0)
 &\approx &
\frac{3}{4\pi}\int_0^{\kf}\!\frac{d^3p}{\ep{p}^2}
\frac{\vct{p}^2-p_y^2}{2\ep{p}} \nonumber \\
&=&
\ekf^2
\left(
\frac32\vf-\vf^3-\frac34\left(1-\vf^2\right)
\log\frac{1+\vf}{1-\vf}
\right) \nonumber \\
&=&
\kf^2\,\frac{\vf^3}{5}\left( 1+\frac37\vf^2+\cdots\right), 
\end{eqnarray*}
where $\vf$ denotes the Fermi velocity $\kf/\ekf$.
Thus,  in contrast to the imaginary part, the real part of 
${\mit\Pi}_{\text{Pauli}}$ and ${\mit\Pi}_{\text{no-sea}}$ may
yield a small contribution to the excitation energy and strength
of the giant GT resonance state in the region $q_0\ll 2M^\ast$.  

The response function $R(q_0)$ is defined with the above
correlation functions as\cite{ks1}
\begin{eqnarray*}
R(q_0)=\frac{1}{\pi}\frac{V}{(2\pi)^3}
 \, \text{Im}\,{\mit\Pi}(q_0),
\end{eqnarray*}
$V$ standing for the volume of the system $(3\pi^2/2\kf^3)A$.
Then, the total GT strength in the mean field approximation
is obtained by integrating it over $q_0$.
The contribution from the particle-hole states to the strength
of the $\beta_-$ transitions is
\begin{eqnarray}
S_{{\text{ph}}}^{(-)} &=&
\frac{1}{\pi}\frac{V}{(2\pi)^3}
%%%%%
\lim_{\eta\rightarrow+0}
 \int_{-\eta}^\infty \!dq_0\,
%%%%%  \int_0^\infty \!dq_0\,
 \text{Im} {\mit\Pi}_{\text{ph}}(q_0) \nonumber  \\
&=& \frac{4V}{(2\pi)^3}
\int\!d^3p\, \left(\thn{p}-\thp{p}\right)\,
\frac{M^{\ast2}+p_y^2}{E_{\vct{p}}^2}, \label{phsum}
\end{eqnarray}
where the variable $\eta$ is introduced, since
$\text{Im} {\mit\Pi}_{\text{ph}}(q_0)$ contains $\delta(q_0)$.
This can be written with use of Eq.(\ref{qkf}) as
\begin{eqnarray*}
S_{{\text{ph}}}^{(-)}=\frac{A}{\kf^3}
\left(Q(\kfn)-Q(\kfp)\right).
\end{eqnarray*}
The function $Q(\kfn)-Q(\kfp)$ can be expanded
in terms of $(\kfn-\kfp)$,
\begin{eqnarray*}
Q(\kfn)-Q(\kfp)\approx \frac{dQ(\kf)}{d\kf}(\kfn-\kfp) + \cdots,
\end{eqnarray*}
where we have
\begin{eqnarray*}
\frac{dQ(\kf)}{d\kf}
= 3\kf^2\left(1-\frac{2}{3}v_{\text{F}}^2\right).
\end{eqnarray*}
When utilizing as usual the relationship:
\begin{eqnarray*}
\kfn-\kfp\approx \frac{2}{3}\kf\frac{N-Z}{A},
\end{eqnarray*}
the GT strength in the nucleon sector is approximately written as
\begin{eqnarray}
S_{{\text{ph}}}^{(-)}
\approx \left(1-\frac{2}{3}v_{\text{F}}^2\right)2(N-Z).
\label{rlsr}
\end{eqnarray}

In the present definition of the spin-isospin operators, the
well-known IFF sum rule is described as
\begin{eqnarray}
\mtrix{}{Q_+Q_-}{}- \mtrix{}{Q_-Q_+}{} = 2(N-Z) \label{sr}
\end{eqnarray}
with
\[
 Q_\pm = \sum_i^A\tau_{\pm i}\sigma_{yi}\,. %\label{sr} 
\]
This sum rule is nothing but a result of the the commutation
relation:\,
$\left[\,\tau_+\sigma_y\,,\,\tau_-\sigma_y\,\right] = 2\tau_z.$
If we assume that $Q_+\ket{\ } = 0$,
then we simply obtain
\begin{eqnarray}
\mtrix{}{Q_+Q_-}{} = 2(N-Z). \label{sr2}
\end{eqnarray}
Comparing Eq.(\ref{rlsr}) with the above equation, it is seen that
the GT strength in the nucleon sector of the relativistic model
is quenched by a factor $(1-\frac{2}{3}v_{\text{F}}^2)$. In fact,
as shown below, the quenched amount is taken by the
nucleon-antinucleon pair excitations. 
In the present model, there is no GT strength of the  $\beta_+$
transitions in the particle-hole sector,
\begin{eqnarray*}
S_{{\text{ph}}}^{(+)}
%%%%%%
=\frac{1}{\pi}\frac{V}{(2\pi)^3}\lim_{\eta\rightarrow+0}
 \int_{\eta}^\infty \!dq_0\,
 \text{Im} {\mit\Pi}_{\text{ph}}(-q_0) 
%%%%%%
=0.
\end{eqnarray*}

The total GT strength of the proton-antineutron($\beta_-$)
excitations in NFA is given by
\begin{eqnarray}
S_{{\text{Pauli}}}^{(-)}&=&\frac{1}{\pi}\frac{V}{(2\pi)^3}
 \int_0^\infty \!dq_0\, \text{Im}{\mit\Pi}_{\text{Pauli}}(q_0)
\nonumber \\
& =&
- \frac{4V}{(2\pi)^3}
\int\!d^3p\, \, 
\thp{p}\,\frac{\vct{p}^2-p_y^2}{E_{\vct{p}}^2}\,,\label{paulisum-}
\end{eqnarray}
which is negative, as mentioned before. The one of the
neutron-antiproton($\beta_+$) excitations is obtained in the same
way by changing the sign of $q_0$ in Eq.(\ref{pi_pauli2}),
\begin{eqnarray}
S_{{\text{Pauli}}}^{(+)}&=&\frac{1}{\pi}\frac{V}{(2\pi)^3}
 \int_0^\infty \!dq_0\,
 \text{Im}{\mit\Pi}_{\text{Pauli}}(-q_0) \nonumber \\
& =&
- \frac{4V}{(2\pi)^3}
\int\!d^3p\, \, 
\thn{p}\,\frac{\vct{p}^2-p_y^2}{E_{\vct{p}}^2}\,.\label{paulisum+}
\end{eqnarray}

From Eqs.(\ref{phsum}), (\ref{paulisum-}) and (\ref{paulisum+}),
we obtain the GT sum value in the mean field approximation of NFA,
\begin{eqnarray}
S_{{\text{ph}}}^{(-)}+S_{{\text{Pauli}}}^{(-)}
 -S_{{\text{Pauli}}}^{(+)}=2(N-Z),\label{rlsumrule}
\end{eqnarray}
which is just IFF sum rule. Thus the quenched amount of the GT
strength in the nucleon sector is taken by the
nucleon-antinucleon pairs, and the Pauli blocking terms are
necessary for NFA to satisfy the sum rule.

In NSA, the GT strength of the particle-hole pairs is the same as
in NFA, but the strengths from the antinucleon degrees of freedom
are different. For the $\beta_-$ transitions, the integration of
Eq.(\ref{pi_nosea2}) over negative excitation energy  gives
\begin{eqnarray*}
S_{{\text{no-sea}}}^{(-)}&=&\frac{1}{\pi}\frac{V}{(2\pi)^3}
 \int_{-\infty}^0 \!dq_0\,
 \text{Im}\,{\mit\Pi}_{\text{no-sea}}(q_0)  \nonumber \\
&=& \frac{4V}{(2\pi)^3}
\int\!d^3p\,
\thn{p}\,\frac{\vct{p}^2-p_y^2}{E_{\vct{p}}^2}\,.
\end{eqnarray*}
This is equal to $-S_{{\text{Pauli}}}^{(+)}$
as seen in Eq.(\ref{paulisum+}),
\begin{eqnarray*}
S_{{\text{no-sea}}}^{(-)}
 =-S_{{\text{Pauli}}}^{(+)}\,.
\end{eqnarray*}
In the same way, we obtain the
relationship:
\begin{eqnarray*}
S_{{\text{no-sea}}}^{(+)}
 = -S_{{\text{Pauli}}}^{(-)},
\end{eqnarray*}
where the l.h.s denotes the total strength of the $\beta_+$
antinucleon-hole excitations in NSA. In NSA, each strength is
positive, but the energy-weighted sum becomes negative.
Although the strengths of the $\beta_-$ and $\beta_+$
transitions in NSA are different from those in NFA, their
difference satisfies the sum rule, as in Eq.(\ref{rlsumrule}),
\begin{eqnarray}
S_{{\text{ph}}}^{(-)}+S_{{\text{no-sea}}}^{(-)}
 -S_{{\text{no-sea}}}^{(+)}=2(N-Z).\label{rlsumrule2}
\end{eqnarray}

\section{The GT Sum Rule in Relativistic RPA}

\label{sec4}

In non-relativistic models, RPA correlations in the GT states
are assumed to be induced by
Landau-Migdal(LM) parameter $g'$\cite{bm,suzuki}.
In the relativistic model, we also introduce $g'$ which is
provided in the nuclear Lagrangian as a contact term with
the pseudo vector coupling:
\begin{eqnarray}
 {\cal L}=
\frac{g_5}{2}\,
 \overline{\psi}\mitg^\mu_i\psi\,
 \overline{\psi}\mitg_{\mu i}\psi\,,\quad
\label{L}
\end{eqnarray}
where
\begin{eqnarray*}
\mitg^\mu_i=\gamma_5\gamma^\mu\tau_i \,, \quad
g_5=\left(\frac{f_\pi}{m_\pi}\right)^2g'.
%%\label{L}
\end{eqnarray*}
Although
it is not unique how to introduce $g'$ in the relativistic
model, we have shown  that the above Lagrangian yields the known
expression for the excitation energy of the GT state in the
non-relativistic limit\cite{ksg1,ksg1-2}. We note that, for
example, the GT state can not be described in relativistic
models by putting the LM parameter 
into the meson propagators\cite{ksg1-2}.

For the Lagrangian Eq.(\ref{L}), the RPA correlation
function ${\mit\Pi}_{\text{RPA}}$ is written in terms of the
mean field one ${\mit\Pi}$ as, 
\begin{eqnarray*}
{\mit\Pi}_{\text{RPA}}(q_0)
 =\frac{{\mit\Pi}(q_0)}{1+\chi_5{\mit\Pi}(q_0)}\,, \quad
\chi_5=
\frac{g_5}{(2\pi)^3}.
\end{eqnarray*}
Then the RPA response functions for the $\beta_-$ and $\beta_+$
transitions are given, respectively, by
\begin{eqnarray*}
R_{\text{RPA}}^{(\mp)}(q_0)=\frac{1}{\pi}\frac{V}{(2\pi)^3}
 \, \text{Im}\,{\mit\Pi}_{\text{RPA}}(\pm q_0).
\end{eqnarray*}
In NFA, ${\mit\Pi}(q_0)$ is provided by Eqs.(\ref{p-h2})
and (\ref{pi_pauli2}) as,
\begin{widetext}
\begin{eqnarray}
{\mit\Pi}(q_0)
 =-\,\frac{\alpha_{\text{ph}}}{q_0+i\varepsilon}
 -4\int\!d^3p\,
 \frac{\vct{p}^2-p_y^2}{E_{\vct{p}}^2}
 \left(
 \frac{\theta_{\vct{p}}^{\text{(p)}}}
 {2E_{\vct{p}}-q_0-i\varepsilon}
 +
 \frac{\theta_{\vct{p}}^{\text{(n)}}}
 {2E_{\vct{p}}+q_0-i\varepsilon}
 \right),\label{nfapi}
\end{eqnarray}
while the one in NSA is given by Eqs.(\ref{p-h2}) and
(\ref{pi_nosea2}) as,
\begin{eqnarray}
 {\mit\Pi}(q_0)
 =-\,\frac{\alpha_{\text{ph}}}{q_0+i\varepsilon}
 -4\int\!d^3p\,
 \frac{\vct{p}^2-p_y^2}{E_{\vct{p}}^2}
 \left(
 \frac{\theta_{\vct{p}}^{\text{(p)}}}
 {2E_{\vct{p}}-q_0+i\varepsilon}
 +
 \frac{\theta_{\vct{p}}^{\text{(n)}}}
 {2E_{\vct{p}}+q_0+i\varepsilon}
 \right).\label{pinsa}
\end{eqnarray}

\subsection{The Non Energy-Weighted GT Sum Rule}

In NFA, the non energy-weighted GT sum value in RPA  is given by
\begin{eqnarray*}
S_{{\text{RPA}}}^{(0)}= 
\lim_{\eta\rightarrow +0}\left\{\int_{-\eta}^\infty dq_0 \,
R_{\text{RPA}}^{(-)}(q_0) - \int_{\eta}^\infty dq_0 \,
R_{\text{RPA}}^{(+)}(q_0)\right\}.
\end{eqnarray*}
In order to calculate the r.h.s.,
we expand ${\mit\Pi}_{\text{RPA}}$ in terms of $\chi_5$,
\begin{eqnarray}
S_{{\text{RPA}}}^{(0)}
=\frac{1}{\pi}\frac{V}{(2\pi)^3}\,\sum_{n=0}^\infty
\left(-\chi_5\right)^n\, \lim_{\eta\rightarrow +0}\,I_n^{(0)},
\label{taylor}
\end{eqnarray}
where $I_n^{(0)}$ is defined as
\begin{eqnarray}
I_n^{(0)}=\int_{-\eta}^\infty\!dq_0\,
 \text{Im}\left( {\mit\Pi}(q_0)\right)^{n+1}
-
\int_{\eta}^\infty\!dq_0\,
\text{Im}\left( {\mit\Pi}(-q_0)\right)^{n+1}.\label{In}
\end{eqnarray}
Since ${\mit\Pi}(\pm q_0)$ has no pole in the first quadrant of
the complex plane, the integration on the closed contour provides
us with\cite{bertsch}
\begin{eqnarray}
\int_{-\eta}^R\!dq_0 \left( {\mit\Pi}(q_0)\right)^{n+1}
-i\int_{0}^R\!dq_0\,\, \left( {\mit\Pi}(iq_0-\eta)\right)^{n+1} 
+\int_C\!dq_0 \left( {\mit\Pi}(q_0)\right)^{n+1}=0, \\
\noalign{\vskip4pt}
\int_{\eta}^R\!dq_0 \left( {\mit\Pi}(-q_0)\right)^{n+1}
-i\int_{0}^R\!dq_0 \left( {\mit\Pi}(-iq_0-\eta)\right)^{n+1} 
+\int_C\!dq_0 \left( {\mit\Pi}(-q_0)\right)^{n+1}=0,
\end{eqnarray}
where $C$ indicates the contour on
$q_0=R\,e^{i\theta}\,, ( 0\le\theta\le \pi/2 )$.
Using the above two equations together with the fact that 
${\mit\Pi}^\ast(-iq_0-\eta)={\mit\Pi}(iq_0-\eta)$,
$I_n$ can be expressed as
\begin{eqnarray}
 I_n^{(0)}=-\,\text{Im}\int_C\!dq_0\,
\left[
 \left( {\mit\Pi}(q_0)\right)^{n+1}
- \left( {\mit\Pi}(-q_0)\right)^{n+1}
 \right].\label{ic}
\end{eqnarray}
When $|q_0|\rightarrow\infty$, ${\mit\Pi}(q_0)$ in Eq.(\ref{nfapi})
behaves as
\begin{eqnarray}
{\mit\Pi}(q_0)
 =-\,\frac{2(2\pi)^3}{q_0}\frac{N-Z}{V}.\label{pi_infty}
\end{eqnarray}
Hence, the integration on the contour $C$ gives
\begin{equation}
 \int_C\!dq_0\,\, \left( {\mit\Pi}(\pm q_0)\right)^{n+1}
 =\frac{i}{R^n}\left(\mp\, 2(2\pi)^3 \frac{N-Z}{V} \right)^{n+1}
 \int_0^{\pi/2}\!d\theta\,e^{-in\theta}\,\rightarrow 0\,,\,\,\,
 (\,R \rightarrow \infty\,,\, n \ge 1\, ).
 \label{eq_int_contour}
\end{equation}
\end{widetext}
From Eqs.(\ref{ic}) and (\ref{eq_int_contour}), finally we obtain
\begin{eqnarray*}
I_n^{(0)}=0\,, \quad  ( n\ge1\,).
\end{eqnarray*}
Hence, the GT sum value of RPA in NFA
is equal to the one in the mean field approximation
Eq.(\ref{rlsumrule}),  
\begin{eqnarray*}
S_{{\text{RPA}}}=\frac{1}{\pi}\frac{V}{(2\pi)^3}\,
 \lim_{\eta\rightarrow +0}\,I^{(0)}_0 = 2(N-Z).
\end{eqnarray*}
Thus, RPA in NFA satisfies IFF sum rule.

We can calculate in the same way the GT sum value of RPA in NSA,
but by defining $J_n^{(0)}$ instead of $I_n^{(0)}$
in Eq.(\ref{taylor}),
\begin{eqnarray*}
J_n^{(0)}&=&\int_{-\infty}^{\eta} \!dq_0\,\text{Im}
 \left( {\mit\Pi}(q_0)\right)^{n+1} \nonumber \\
& & -
 \int_{-\infty}^{-\eta} \!dq_0\,\text{Im}
 \left( {\mit\Pi}(-q_0)\right)^{n+1},
\end{eqnarray*}
where ${\mit\Pi}(q_0)$ is given by Eqs.(\ref{pinsa}).
By taking a closed contour with no pole in the second quadrant
in this case, we can prove that $J_n^{(0)}=0\, (n\ge1)$, and
therefore, NSA also satisfies IFF sum rule.

In the above, we have shown that both NFA and NSA satisfy
IFF sum rule. This means that the difference between
total GT strengths for the $\beta_-$ and $\beta_+$ transitions is
independent of the value of LM parameter $g'$. 
Of course, however, the strength of the giant GT resonance state
and each total strength of the $\beta_-$ and $\beta_+$ transitions
depend on the value of the parameter.  
Let us estimate each strength separately.

We divide the excitation-energy
region into two parts:\,$|q_0|<2M^\ast$ and $|q_0|\geq 2M^\ast$.
In the first region
$|q_0|<2M^\ast$, we have a discrete state which corresponds to the
giant GT resonance state in both NFA and NSA. In this region, we
have ${\mit\Pi}_{\text{Pauli}}^{\text{(I)}}
= {\mit\Pi}_{\text{no-sea}}^{\text{(I)}}=0$,
so that the RPA response function
is described as
\begin{eqnarray*}
R_{\text{RPA}}(q_0)=
-\,\frac{1}{\chi_5}\frac{V}{8\pi^4}
 \, \text{Im}\frac{q_0}{F_{\text{T}}(q_0)+i\varepsilon}
\end{eqnarray*}
with
\begin{eqnarray*}
F_{\text{T}}(q_0)=q_0+\chi_5\left( -\,\alpha_{\text{ph}}
 +q_0{\mit\Pi}_{\text{Pauli}}^{\text{(R)}}(q_0) \right).
\end{eqnarray*}
In the case of NSA, ${\mit\Pi}_{\text{Pauli}}^{\text{(R)}}(q_0)$
is replaced with ${\mit\Pi}_{\text{no-sea}}^{\text{(R)}}(q_0)$,
but they are the same, as mentioned before.
The excitation energy $\omega_0$ of the giant GT resonance state
is given by the solution of the equation:
\begin{eqnarray*}
F_{\text{T}}(\omega_0)=0 . 
\end{eqnarray*}
When we expand $F_{\text{T}}(q_0)$ at $\omega_0$ as,
\begin{eqnarray*}
 F_{\text{T}}(q_0)
 =F'_{\text{T}}(\omega_0)(q_0-\omega_0)+O((q_0-\omega_0)^2),
\end{eqnarray*}
we can describe the response function in the form:
\begin{eqnarray*}
R_{\text{RPA}}(q_0)
=\frac{1}{\chi_5}\frac{V}{8\pi^3}
\frac{\omega_0}{F'_{\text{T}}(\omega_0)}
\,\delta(q_0-\omega_0).
\end{eqnarray*}
From this equation, we obtain the excitation strength of
the GT state:
\begin{eqnarray}
S_{\text{GT}}
&=&\frac{1}{\chi_5}\frac{V}{8\pi^3}
\frac{1}{D'_{\text{T}}(\omega_0)}\,, \\
\label{sgt}
D'_{\text{T}}(\omega_0)
&=&\chi_5\left( \frac{\alpha_{\text{ph}}}{\omega_0^2}
 + \frac{d}{d\omega_0}
 {\mit\Pi}_{\text{Pauli}}^{\text{(R)}}(\omega_0) \right),
\nonumber
\end{eqnarray}
using the dimesic function $D_{\text{T}}(q_0)$ defined by
\[
F_{\text{T}}(q_0)=q_0D_{\text{T}}(q_0).
\]
If we neglect ${\mit\Pi}_{\text{Pauli}}^{\text{(R)}}$, which is
small at $|q_0|\ll 2M^\ast $, $\omega_0$ and $S_{\text{GT}}$
are approximately given by
\begin{equation}
\omega_0\approx \chi_5 \alpha_{\text{ph}}\,,\qquad
S_{\text{GT}}\approx  \frac{V}{(2\pi)^3}\alpha_{\text{ph}}.
\label{ex}
\end{equation}
Thus, the value of $S_{\text{GT}}$ is almost independent of the
value of the LM parameter.

\begin{widetext}
In the region $|q_0|\geq 2M^\ast$, the RPA response function
in NFA is described as
\begin{eqnarray*}
R_{\text{RPA}}(q_0)
 &=&
\frac{V}{8\pi^4}
\frac{{\mit\Pi}_{\text{Pauli}}^{\text{(I)}}(q_0)}
{
\left(1+\chi_5{\mit\Pi}_r(q_0)\right)^2
+\left(\chi_5{\mit\Pi}_{\text{Pauli}}^{\text{(I)}}(q_0)\right)^2
}\,,\qquad
{\mit\Pi}_r(q_0)=-\,\frac{\alpha_{\text{ph}}}{q_0}
+{\mit\Pi}_{\text{Pauli}}^{\text{(R)}}(q_0).
\end{eqnarray*}
This equation provides us with 
the total GT strength of the $\beta_-$ transitions:
\begin{eqnarray}
S_{\text{NFA}}^{(-)}&=&
\int_{2M^\ast}\!dq_0\,R_{\text{RPA}}(q_0)=
\frac{V}{8\pi^4}
\int_{2M^\ast}^{2\ekfp}\!dq_0\,
\frac{I_{\text{Pauli}}(k_p,q_0)}
{\left(1+\chi_5{\mit\Pi}_r(q_0)\right)^2
+\left(\chi_5I_{\text{Pauli}}(k_p,q_0)\right)^2}\,,\label{spauli-}
\end{eqnarray}
and that of the $\beta_+$ transitions:
\begin{eqnarray}
S_{\text{NFA}}^{(+)}=
\int_{2M^\ast}\!dq_0\,R_{\text{RPA}}(-\,q_0)
&=&
\frac{V}{8\pi^4}
\int_{2M^\ast}^{2\ekfn}\!dq_0\,
\frac{I_{\text{Pauli}}(k_n,q_0)}
{\left(1+\chi_5{\mit\Pi}_r(-q_0)\right)^2
+\left(\chi_5I_{\text{Pauli}}(k_n,q_0)\right)^2}\,,\label{spauli+}
\end{eqnarray}
\end{widetext} 
where we have used the notations:
\[
\ekfi=\sqrt{M^{\ast2}+k_i^2}\,,\quad (\ \text{i}=\text{p, n}\ ).
\]
The bounds of integral are determined
by Eq.(\ref{Ipauli}). 

The GT strengths of the region $|q_0|\geq 2M^\ast$ in NSA are also
calculated in the same way.
They are obtained in terms of those in NFA as,
\begin{eqnarray}
S_{\mbox{\scriptsize NSA}}^{(-)}=-\,S_{\text{NFA}}^{(+)}\,,\qquad
S_{\mbox{\scriptsize NSA}}^{(+)}=-\,S_{\text{NFA}}^{(-)}.
\label{nfa-nsa}
\end{eqnarray}
Thus, each total strength of the $\beta_-$ and $\beta_+$
transitions in NSA is different from the one in NFA.
We note that the above relationships are obtained by performing
the integration of the NSA response functions 
over negative energy, as we did before.

%%%%%%%%%
%  table 

\begin{table}%[h]
 
\caption{Dependence of the excitation energy $\omega_0$ and 
strength $S_{{\text{GT}}}$ of the GT state on Landau-Migdal
parameter  $g'$. For details, see the text.}

\begin{center}
\begin{tabular}{|c|c|ccc|c|} \hline
 \smash{\lower1.5ex\hbox{$g'$}} 
 & $\omega_0$  & $S_{\text{GT}}/V$&
 $S_{\text{Pauli}}^{(-)}/V$ & $S_{\text{Pauli}}^{(+)}/V$ &
 $S_{\text{total}}/V$ \\[2pt]
   & MeV & fm$^{-3}$ & fm$^{-3}$ & fm$^{-3}$ & fm$^{-3}$ \\[2pt]
 \hline
 $0.0$ & $ 0.0\ \  $    & $ 0.06072$ & $-0.00733$ & $-0.01524$
 & $0.06863$\\
 $0.5$ & $ 11.944$ & $ 0.06116$ & $-0.00750$ & $-0.01497$
 & $0.06863$\\
 $1.0$ & $ 23.975$ & $ 0.06161$ & $-0.00767$ & $-0.01470$
 & $0.06863$\\
 $2.0$ & $ 48.300$ & $ 0.06250$ & $-0.00804$ & $-0.01417$
 & $0.06863$\\
 $3.0$ & $ 72.977$ & $ 0.06341$ & $-0.00844$ & $-0.01366$
 & $0.06863$\\
 $4.0$ & $ 98.012$ & $ 0.06432$ & $-0.00887$ & $-0.01318$
 & $0.06863$\\
\hline
\end{tabular} 
\end{center}
\label{table} 
\end{table}
%%%%%%%%%

In Table \ref{table}, we list the excitation energy and strength
of the giant GT resonance state as a function of $g'$, and
compare the strength of each energy region
with the total RPA strength in NFA:
\begin{eqnarray*}
S_{\text{total}}
=S_{\text{GT}}+S_{\text{NFA}}^{(-)}-S_{\text{NFA}}^{(+)}.
\end{eqnarray*}
In addition to $(f_\pi/m_\pi)^2=392$MeV$\cdot$fm$^3$, 
we have employed the values of
parameters, as an example,  $M^\ast/M=0.6$\,, \
$k_n=1.4\mbox{\,fm}^{-1}$
and $k_p=1.2\mbox{\,fm}^{-1}$. These give the sum rule
value in the unit of the nuclear volume as
\begin{eqnarray}
2\frac{N-Z}{V}=\frac{2}{3\pi^2}\left( k_n^3-k_p^3\right)
=0.06863\ \mbox{fm}^{-3}\,.\label{ikeda}
\end{eqnarray}
It is seen in the last column in Table \ref{table} that this value
is reproduced in NFA.
Table \ref{table} also shows that the strength of the giant GT
resonance state does not depend strongly on the value of $g'$,
as expected from the previous discussions.
Comparing with sum rule value,
$S_{{\text{GT}}}$ is quenched by 12\% $\sim$ 10\% for $g'= 0\sim 1$.
From the experimental value for the excitation energy of the giant
GT resonance state, $g'$ is estimated to be about 0.6\cite{sakai}. 
In the present case, the approximate values given by Eq.(\ref{ex})
are
\begin{eqnarray*}
 \omega_0\approx 23.801\,g'\,\mbox{MeV}\,,\qquad
 \frac{S_{\text{GT}}}{V}\approx 0.06072\,\mbox{fm}^{-3}.
\end{eqnarray*}

\subsection{The Energy-Weighted GT Sum Rule}

In no charge-exchange excitations, there is a famous theorem on the
energy-weighted sum value of the excitation strengths in RPA by
Thouless\cite{thou}. According to the RPA theorem, the sum value
is equal to the expectation value of the double commutator of
the nuclear Hamiltonian with the excitation operator.
Here the expectation value is calculated by
the ground state in the Hartree-Fock approximation. If the double
commutator is a c-number, the sum value is given by the
model-independent sum rule. Even if it is not a c-number,
the theorem has been frequently used for a justification of the
numerical results.
In this subsection, we will show that the same theorem holds for
charge-exchange excitations with respect to the sum value 
of the $\beta_-$ and $\beta_+$ transition strengths.  

In the case of NFA, the energy-weighted GT sum value is given by 
\begin{eqnarray*}
S_{{\text{RPA}}}^{(1)}&=&\frac{1}{\pi}\frac{V}{(2\pi)^3}
 \lim_{\eta\rightarrow 0}
 \left(
 \int_{-\eta}^\infty\!\!dq_0\,
 q_0\,\text{Im}\,{\mit \Pi}_{\text{RPA}}(q_0)
 \right. \nonumber \\
& &
\phantom{\frac{1}{\pi}\frac{V}{(2\pi)^3}} 
 \left.
+ \int_\eta^\infty\!\!dq_0\,
q_0\,\text{Im}\,{\mit \Pi}_{\text{RPA}}(-q_0)
\right),
\end{eqnarray*}
where the first term of the r.h.s. stands for the energy-weighted
sum of the $\beta_-$ transition strengths, while the second term
that of the $\beta_+$ ones.
In the same way as in the previous subsection, we expand
${\mit \Pi}_{\text{RPA}}$ in terms of the coupling constant
$\chi_5$,
\begin{eqnarray}
S_{{\text{RPA}}}^{(1)}
 =\frac{1}{\pi}\frac{V}{(2\pi)^3}\,\sum_{n=0}^\infty
 \left(-\chi_5\right)^n\, \lim_{\eta\rightarrow +0}\,I_n^{(1)},
 \label{taylor1}
\end{eqnarray}
where $I_n^{(1)}$ is defined as
\begin{eqnarray}
I_n^{(1)}&=&\int_{-\eta}^\infty\!dq_0\,q_0\text{Im}
 \left( {\mit\Pi}(q_0)\right)^{n+1} \nonumber  \\
& &
 +
 \int_{\eta}^\infty\!dq_0\,q_0\text{Im}
 \left( {\mit\Pi}(-q_0)\right)^{n+1}.\label{In1}
\end{eqnarray}
From the same arguments for Eq.(\ref{ic})
the above equation can be expressed as 
\begin{eqnarray*}
 I^{(1)}_n=-\,\text{Im}\int_C\!dq_0\,q_0
\left[
 \left( {\mit\Pi}(q_0)\right)^{n+1}
+ \left( {\mit\Pi}(-q_0)\right)^{n+1}
 \right].
\end{eqnarray*}
At $|q_0|\rightarrow\infty$, the integration on the contour $C$
gives
\begin{eqnarray}
& &\int_C\!dq_0\,q_0\, \left( {\mit\Pi}(\pm q_0)\right)^{n+1}
\nonumber \\ 
&=&\frac{i}{R^{n-1}}
 \left(\mp\, 2(2\pi)^3 \frac{N-Z}{V} \right)^{n+1}
\!\!\int_0^{\pi/2}\!\!\!d\theta\,e^{-i(n-1)\theta}.\ \ \ \ \ 
 \label{eq_int_contour1}
\end{eqnarray}
Consequently, we obtain
\begin{eqnarray*}
 I_n^{(1)}=0\,, \,\,\, (\, n\ge 2\,).
\end{eqnarray*}
Thus the energy-weighted sum value is given by the two terms,
\begin{eqnarray*}
 S_{{\text{RPA}}}^{(1)} =
\frac{1}{\pi}\frac{V}{(2\pi)^3}\left(
 I_0^{(1)} -\chi_5 I_1^{(1)}
  \right),
\end{eqnarray*}
where $I_0^{(1)}$ is calculated using Eq.(\ref{nfapi}),
\begin{equation}
 I_0^{(1)}=-\,8\pi\int\!\!d^3p\,
 \frac{\vct{p}^2-p_y^2}{E_{\vct{p}}}\,
 \left(
 \theta_{\vct{p}}^{\text{(p)}}
 +\theta_{\vct{p}}^{\text{(n)}}
 \right), \label{eq_ew_0}
\end{equation}
while $I_1^{(1)}$ is obtained directly from
Eq.(\ref{eq_int_contour1}),
\begin{eqnarray}
I_1^{(1)}
&=&
-\,\pi
\left(
2(2\pi)^3
\frac{N-Z}{V}
\right)^2.
\end{eqnarray}
We note that Eq.(\ref{eq_int_contour1}) should not be used for
$I_0^{(1)}$ which requires the $1/q_0^2$ term in
Eq.(\ref{pi_infty}). Finally, we obtain the energy-weighted sum
of the $\beta_-$ and $\beta_+$ transition strengths in RPA of NFA,
\begin{eqnarray}
 S_{{\text{RPA}}}^{(1)}
&= &-\,\frac{V}{\pi^3}
 \int\!\!d^3p\,
 \frac{\vct{p}^2-p_y^2}{E_{\vct{p}}}\,
\left(
 \theta_{\vct{p}}^{\text{(n)}}
+ \theta_{\vct{p}}^{\text{(p)}}
 \right) \nonumber \\
& &
 +4g_5\frac{(N-Z)^2}{V}. \label{eq_e_sum}
\end{eqnarray}

If the antinucleon degrees of freedom were neglected, we would
have $I_0^{(1)}=0$ and $I_1^{(1)}=-\pi\alpha_{{\text{ph}}}^2$,
which give
\begin{eqnarray}
S_{{\text{RPA}}}^{(1)}
 =\frac{V}{(2\pi)^3}\chi_5\alpha_{\text{ph}}^2.\label{sgtrpa1}
\end{eqnarray}
This is equal to $\omega_0S_{{\text{GT}}}$ from Eqs.(\ref{ex}).
In fact, even if we take into account
the coupling between the particle-hole states and the
nucleon-antinucleon states, the energy-weighted
strength of the giant GT resonance state is approximately given
by the above equation,
\begin{eqnarray}
\omega_0S_{{\text{GT}}}
& \approx& \frac{V}{(2\pi)^3}\chi_5\alpha_{\text{ph}}^2
\nonumber \\
& \approx& \left(1-\frac23\vf^2\right)^2 4g_5
       \frac{\left(N-Z\right)^2}{V},\label{sgtrpa2}
\end{eqnarray}
since the coupling is weak. This implies that
the energy-weighted sum of the strengths in the giant GT
resonance region is quenched by a factor $(1-2\vf^2/3)^2$,
compared with the non-relativistic one. Its value is about 0.77
for $\vf=0.43$ obtained from $M^\ast=0.6M$ and
$\kf=1.36$fm$^{-1}$.

It is worthwhile noting that we can do formally the above
calculations including the divergent terms
with $G_{\text{F}}\,G_{\text{F}}$.
In this case, $\theta_{\vct{p}}^{(i)}$ is
replaced with $\theta_{\vct{p}}^{(\tau)}-1$ in
Eq.(\ref{eq_ew_0}), so that we have 
\begin{eqnarray}
S_{{\text{RPA}}}^{(1)}
&=&\frac{V}{\pi^3}
\int\!\!d^3p\,
\frac{\vct{p}^2-p_y^2}{E_{\vct{p}}}\,
\left(
2-\theta_{\vct{p}}^{\text{(n)}}
- \theta_{\vct{p}}^{\text{(p)}}
\right) \nonumber \\
& &
+4g_5\frac{(N-Z)^2}{V}. \label{eq_ew_full}
\end{eqnarray}
We can see that the negative contribution in Eq.(\ref{eq_e_sum})
stems from neglect of the divergent term. A part of contribution
from the Pauli blocking terms is included in the second
term of Eq.(\ref{eq_e_sum}) also, as seen from
Eqs.(\ref{sgtrpa1}) and (\ref{sgtrpa2}).

In NSA, we can perform the same calculation, but by defining
$J_n^{(1)}$ instead of $I_n^{(1)}$ in Eq.(\ref{In1}) for NFA,
\begin{eqnarray*}
J_n^{(1)}& =& \int_{-\infty}^{\eta} \!dq_0\,q_0\text{Im}\,
 \left( {\mit\Pi}(q_0)\right)^{n+1} \nonumber \\
& &
 +
 \int_{-\infty}^{-\eta} \!dq_0\,q_0\text{Im}\,
 \left( {\mit\Pi}(-q_0)\right)^{n+1},
\end{eqnarray*}
with ${\mit\Pi}(q_0)$ given in Eq.(\ref{pinsa}).
We can show that the energy-weighted sum in NSA is
the same as Eq.(\ref{eq_e_sum}) for NFA.

Now we will show that Eq.(\ref{eq_e_sum}) is equal to the
expectation value of the double commutator of the Hamiltonian
with the GT operator, when the expectation value is calculated
by the ground state in the mean field.
The nuclear field in the present model is written as
\begin{eqnarray}
 \psi(\vct{x})&=&\int\!\frac{d^3p}{(2\pi)^{3/2}}\sum_{\alpha}
\left(
u_\alpha(\vct{p})\exp(i\vct{p}\cdot\vct{x})\,a_\alpha(\vct{p})
\right. \nonumber \\
& &
\left. 
+v_\alpha(\vct{p})\exp(-i\vct{p}\cdot\vct{x})\,
b^\dagger_\alpha(\vct{p})
\right),\label{field}
\end{eqnarray}
where the first term in the parentheses
describes the particles and the second the antiparticles.
The suffix 
$\alpha$ denotes the spin and isospin quantum numbers as 
$u_\alpha(\vct{p})=u_\sigma(\vct{p})\,\ket{\tau}$\,,
( $\alpha=\sigma, \tau$ ), and the positive and negative spinors
are given by
\begin{eqnarray*}
u_\sigma(\vct{p})&=&\left[\frac{E_p+M^{\ast}}{2E_p}\right]^{1/2}
\left(
\begin{array}{cc}
1\\
\displaystyle{\frac{\gvct{\sigma}\cdot\vct{p}}{E_p+M^{\ast}}}
\end{array}
\right)\xi\,, \\
\noalign{\vskip4pt} 
v_\sigma(\vct{p})&=&
\left[\frac{E_p+M^{\ast}}{2E_p}\right]^{1/2}
\left(
\begin{array}{cc}
\displaystyle{\frac{\gvct{\sigma}\cdot\vct{p}}{E_p+M^{\ast}}}\\
1							      
\end{array}
\right)\xi\,
\end{eqnarray*}
with the Pauli spinor $\xi$. The creation and annihilation
operators, $a_\alpha(\vct{p})$ and $b^\dagger_\alpha(\vct{p})$,
are defined as usual. Then, the mean field Hamiltonian, the
interaction and the GT field
operator are described, respectively, as
\begin{eqnarray*}
H_0&=& \int\!d^3x\,
 \overline{\psi}(\vct{x})
 \left(-i\,\gvct{\gamma}\cdot\gvct{\nabla} +M^\ast
 \right)\psi(\vct{x})\,,\\
\noalign{\vskip4pt} 
V&=& -\frac{g_5}{2}
 \int\!d^3x\,
 \overline{\psi}(\vct{x})\mitg^\mu_i\psi(\vct{x})\,
 \overline{\psi}(\vct{x})\mitg_{\mu i}\psi(\vct{x})\,, \\
\noalign{\vskip4pt}
F_\pm&=&\int\!d^3x\,
 \overline{\psi}(\vct{x})\gamma_5 \gamma_y \tau_\pm\psi(\vct{x}).
\end{eqnarray*}
We are assuming throughout the present paper that Dirac particles
are bound in the Lorentz scalar and vector potentials, but in the
above mean field Hamiltonian, we have deleted the Lorentz vector
potential which does not appear explicitly in the present
discussions, while the Lorentz scalar
potential is included in the nucleon effective mass $M^*$.

\begin{widetext}

The double commutator of the mean field Hamiltonian with
the GT operator is easily calculated.
By using the relationship for arbitrary operators $A$ and $B$,
\begin{eqnarray*}
 \left[\,\psi^\dagger(\vct{x})A\psi(\vct{x})\,,\,
 \psi^\dagger(\vct{y})B\psi(\vct{y})
\,\right]
=\psi^\dagger(\vct{x})\left[\,A\,,\,B\,\right]\psi(\vct{x})
\,\delta(\vct{x}-\vct{y}),
\end{eqnarray*}
we obtain
\begin{eqnarray*}
\left[\,F_+\,,\,\left[\,H_0\,,\,F_-\, \right]\,\right]
&=&
4i \int\!d^3x\,
 \psi^\dagger(\vct{x})
 \gamma_0
 \left(
\gvct{\gamma}\cdot\gvct{\nabla} -\gamma_y\frac{\partial}{\partial y}
 \right)
\psi(\vct{x}).
\end{eqnarray*}
Its expectation value of the ground state in the mean field becomes
\begin{eqnarray}
\mtrix{ }{%
\left[\,F_+\,,\,\left[\,H_0\,,\,F_-\, \right]\,\right]
}{ }
&=&
-\,\frac{4V}{(2\pi)^3}
\int\!d^3p\,\left[
\left( \theta^{\text{(p)}}_{\vct{p}}
+ \theta^{\text{(n)}}_{\vct{p}}\right)
\,\text{Tr}_{\sigma}
\left(
\left(
 \gvct{\gamma}\cdot\vct{p}-\gamma_y p_y
 \right){\mit\Lambda}_+(\vct{p})
\right) \right. \nonumber \\
& &
\phantom{-\,\frac{4V}{(2\pi)^3}\int\!d^3pp^k}
\left.
+\,2\text{Tr}_{\sigma}\left(
\left(
\gvct{\gamma}\cdot\vct{p}-\gamma_y p_y
\right){\mit\Lambda}_-(-\vct{p})
\right)\right] \nonumber \\
\noalign{\vskip4pt}
&=&
\frac{V}{\pi^3}
\int\!d^3p\,
\frac{\vct{p}^2-p_y^2}{E_{\vct{p}}}
\left(2 -\theta^{\text{(p)}}_{\vct{p}}
- \theta^{\text{(n)}}_{\vct{p}}\right).
\label{eq_ew_rpa1}
\end{eqnarray}

The double commutator of $V$ with the GT operator is calculated in
the same way.
Using the abbreviations:
\begin{eqnarray*}
f^\mu_q=\gamma_0\gamma_5\gamma^\mu\tau_q\,,\quad
X^\mu_q=\left[\,f^\mu_q\,,\,f^2_-\, \right]\,,
\quad (\,q=\pm\,,0\,),
\end{eqnarray*}
it is described as
\begin{eqnarray*}
\left[\,F_+\,,\,\left[\,V\,,\,F_-\, \right]\,\right]
&=&
-\,\frac{g_5}{2}\int\!d^3x\,\left(
\psi^\dagger (X_{\mu\,q})^\dagger\psi\,\,
\psi^\dagger X^\mu_q\psi
+
\psi^\dagger f_{\mu\,q}\psi \,\,
\psi^\dagger\!\left[\,f^2_+\,,\,X^\mu_{-q}\,\right]\!\psi
\right. \nonumber \\
& &
\phantom{-\,\frac{g_5}{2}\int\!d^3x}
\left.
+\,
\psi^\dagger\!\left[\,f^2_+\,,\,X^\mu_{-q}\,\right]\psi\,\,
\psi^\dagger f_{\mu\,q}\psi 
+
\psi^\dagger X^\mu_q \psi \,\,
\psi^\dagger (X_{\mu\,q})^\dagger \psi \right).
\end{eqnarray*}
In the mean field approximation, the expectation value is
calculated neglecting the exchange terms of the matrix elements.
Keeping the only direct terms, we have
\begin{eqnarray*}
\mtrix{}{%
 \left[\,F_+\,,\,\left[\,V\,,\,F_-\, \right]\,\right]
}{}
&=&
-\,\frac{g_5}{2}\int\!d^3x\,\left(
\mtrix{}{\psi^\dagger (X_{\mu\,+})^\dagger\psi\,\,
\psi^\dagger X^\mu_+\psi}{}
+ \mtrix{}{\psi^\dagger X^\mu_+ \psi \,\,
\psi^\dagger (X_{\mu\,+})^\dagger \psi}{} \right. \nonumber\\
 & &
\phantom{-\,\frac{g_5}{2}}
%\int\!d^3}
\left.
+\,
\mtrix{}{\psi^\dagger f_{\mu0}\psi \,\,
\psi^\dagger\!\left[\,f^2_+\,,\,X^\mu_0\,\right]\!\psi}{}
+
\mtrix{}{\psi^\dagger\!\left[\,f^2_+\,,\,X^\mu_0\,\right]\psi\,\,
\psi^\dagger f_{\mu0}\psi}{} \right).
\end{eqnarray*}
The straightforward calculation of the above matrix elements
yields
\begin{eqnarray}
\mtrix{}{%
 \left[\,F_+\,,\,\left[\,V\,,\,F_-\, \right]\,\right]
}{}
&=&
\,g_5 \int\!d^3x\,
\mtrix{}{\psi^\dagger (X^2_+)^\dagger\psi\,\,
\psi^\dagger X^2_+\psi}{} \nonumber \\
\noalign{\vskip4pt}
&=&
g_5\frac{V}{(2\pi)^6}
\left(4\int\!d^3p\,
 \left(
  \theta^{\text{(p)}}_{\vct{p}}-\theta^{\text{(n)}}_{\vct{p}}
 \right)
\right)^2
=4g_5\frac{\left(N-Z\right)^2}{V}.
\label{eq_ew_rpa2}
\end{eqnarray}
\end{widetext} 
Thus, the sum of Eqs.(\ref{eq_ew_rpa1}) and (\ref{eq_ew_rpa2}) is
just equal to the energy-weighted sum of the strengths
for the $\beta_-$ and $\beta_+$ transitions in RPA given in
Eq.(\ref{eq_ew_full}),
\begin{eqnarray*}
 S_{{\text{RPA}}}^{(1)}
= \mtrix{}{[\,F_+\,,\,[\,H_0+V\,,\,F_-\,]\,]}{},
\end{eqnarray*}
including the divergent term.
In Eq.(\ref{eq_e_sum}) for NFA, the divergent term has been
simply neglected.

In NSA, the nuclear field is given by replacing the creation
operator of the antiparticles with the annihilation one in
Eq.(\ref{field})
for NFA. Because of this change, on the one hand, we obtain,
instead of Eq.(\ref{eq_ew_rpa1}),
\begin{eqnarray*}
& &\mtrix{ }{%
\left[\,F_-\,,\,\left[\,H_0\,,\,F_+\, \right]\,\right]
}{ } \nonumber \\
&=&
- \,\frac{V}{\pi^3}
\int\!d^3p\,
\frac{\vct{p}^2-p_y^2}{E_{\vct{p}}}
\left(\theta^{\text{(p)}}_{\vct{p}}
+ \theta^{\text{(n)}}_{\vct{p}}\right),
\end{eqnarray*}
which does not contain the divergent term.
On the other hand, the expectation value of the double commutator
as for
$V$ is the same as Eq.(\ref{eq_ew_rpa2}). Thus, in NFA also, the
energy-weighted sum of the GT strengths in RPA is equal to the
expectation value of the double commutator, when the expectation
value is calculated with the ground state in the mean field.  

Formally we have proved that the RPA theorem holds in
charge-exchange
excitations also, but with respect to the sum of the strengths
for the $\beta_-$ and $\beta_+$ transitions. In the present
relativistic model, however, we have also shown that the
energy-weighted sum value itself 
is divergent. If we neglect simply the divergent terms
as in NFA and NSA, the sum value becomes negative, owing to the
strengths of the Pauli blocking terms or the antinucleon-hole
excitations.
Generally speaking,
all previous calculations in NFA and NSA 
have the same problem.
In order to solve this problem, we need definitely the
renormalization of the divergence.

\section{Discussions and Conclusions}

\label{sec5}

The relativistic model has been extensively used as a
phenomenological
model of nuclei for the past 30 years\cite{sw,ring}.
In particular,
it explains very well the ground-state properties of nuclei
with the mean field approximation, where
antinucleon degrees of freedom are neglected\cite{sw, ring}.
In RPA based on the mean field approximation, however, 
it is known that we can not describe the excited states
within the nucleon space only, and should include at least
Pauli blocking terms from nucleon-antinucleon excitations
in the correlation functions. We have called  this approximation
NFA. The Pauli blocking terms are required for RPA  
to keep  the continuity equation\cite{chin,ks1},
and to reproduce the correct Landau-Migdal(LM) parameters,
etc.\cite{ks2}.
The reason why we need the Pauli blocking terms may be partially
because of the fact that in relativistic models, the complete set 
needs antinucleon degrees of freedom.

In this paper, we have shown that the Pauli blocking terms are
also necessary for RPA to satisfy the Gamow-Teller(GT) sum rule
with respect to the difference between the strengths
of the $\beta_-$ and $\beta_+$ transitions which is
called Ikeda-Fujii-Fujita(IFF) sum rule\cite{iff}. This fact has
been shown
in an analytic way for nuclear matter. If the configuration space is
limited to the nucleon one, the sum rule value is exhausted only by
about 88\%. When adding the Pauli blocking terms to the RPA
correlation function, the sum rule value is reproduced.
The coupling of the particle-hole states with the
nucleon-antinucleon states is weak for reasonable values of the LM
parameter $g'$. As a result, the GT strength which is distributed
over the giant GT resonance region remains to be quenched by about
10 to 12\%.

In the previous paper\cite{ksg1-2}, the GT
strength in the nucleon sector was estimated for finite nuclei
in the mean field approximation. The value of the total strength
was about 94\% of the sum
rule value. The reduction of the quenching is due to
a larger value of the effective mass in finite nuclei than in
nuclear matter. Since the coupling between the
particle-hole and particle-antiparticle states is weak,
the about 6\% quenching of the GT strength is also expected
in RPA of NFA for finite nuclei.
  
The IFF sum rule in the no-sea approximation(NSA)\cite{furun}
has been also investigated.
In NSA, the Dirac sea is assumed to be empty, and the antinucleon
states are treated as particle states with negative energy.
In this way, one can avoid the divergence problem without
violating the continuity equation as in NFA. We have shown that
NSA also satisfies IFF sum
rule, and predicts the same quenching of the strength for the
giant GT resonance state. It has been shown, however,
that each strength of the total $\beta_-$ and $\beta_+$ transitions
is different from those in NFA.

The energy-weighted sum of the GT strengths in RPA is also studied
in NFA and NSA. It has been shown that the sum of the
energy-weighted strengths for the $\beta_-$ and $\beta_+$
transitions is
equal to the expectation value of the double commutator of the
Hamiltonian with the GT operator, when the expectation value is
calculated with the ground state in the mean field and the
divergent terms are deleted. Thus the well-known
RPA theorem by Thouless\cite{thou} holds for charge-exchange
excitations also.
    
We should note finally that renormalization of the divergence
should be investigated in the future study of relativistic models.
So far most of the nuclear observables are well reproduced
phenomenologically without the renormalization. 
NFA and NSA, which satisfy various conservation laws,
are such examples. As shown in the present paper, however,
NFA and NSA provide us with unphysical results also,
as a price of neglecting the divergence.
In the case of the GT excitations, the non-energy
weighted strengths themselves in NFA and the energy-weighted
strengths in NSA  are negative outside of the giant resonance
region. All previous calculations using NFA and NSA 
may have the same problems.

It may depends on nuclear observables whether or not
effects of the renormalization are important. In the previous
studies\cite{ks2}, on the one hand, it was shown that
the LM parameter $F_1$ depends on
antinucleon degrees of freedom, only
through the Pauli blocking terms. These facts were shown by 
renormalized calculations in the $\sigma-\omega$ model.
As a result, some physical quantities 
which are dominated by $F_1$ can be described using 
approximation with the Pauli blocking terms.
Indeed, for example, as well as the center of mass motion,
the space part of the nuclear current, which is responsible
for the orbital part of the nuclear magnetic moments 
was shown to be described well in NFA and NSA\cite{furun,nks}.
On the other hand, effects of the antinucleon degrees of freedom
on the LM parameter $F_0$ are not represented by the Pauli
blocking terms
only, but the contribution from other nucleon-antinucleon
excitations is more important, as shown in ref.\cite{ks2,ks3}.
Another example is the Coulomb sum rule, where the renormalization
provides us with a strong quenching of the sum rule value at high
momentum transfer\cite{ks3}.    
 
Thus, there may be some cases
where the renormalization is not essential for description of the
observables, but generally speaking, we should investigate in the
future how effects of the renormalization change previous
results in relativistic models.  
This may be also true for the GT strengths discussed here, although
the coupling between the particle-hole terms and the Pauli blocking
terms is weak in RPA.

\end{document}